\documentclass[11pt]{article}
\pdfoutput=1

\usepackage{graphics, color,soul}
\usepackage{graphicx}
\usepackage{amssymb}

\usepackage{lmodern,mathtools}

\usepackage{booktabs}
\usepackage[english]{babel}
\usepackage{amsmath,amssymb,amsbsy,amstext, amsthm, simplewick}
\usepackage{hyperref}
\usepackage{tikz}

\usetikzlibrary{decorations.pathmorphing,shapes.misc}
\tikzset{snake it/.style={decorate, decoration=snake}}
\tikzset{cross/.style={cross out, draw=black, minimum size=2*(#1-\pgflinewidth), inner sep=0pt, outer sep=0pt},
cross/.default={1pt}}

\usepackage{amsfonts}
\usepackage{amssymb}
\usepackage{upgreek}
\usepackage{simplewick}
 \usepackage{exscale,relsize}
\usepackage{mathtools}
\usepackage{comment}

\usepackage[margin=1cm,labelfont={sf,bf,scriptsize},textfont={sf,scriptsize}]{caption}

\usepackage{colortbl}
\definecolor{lightgreen}{cmyk}{0.2, 0, 0.2, 0.2}
\definecolor{lightgray}{cmyk}{0.1,0.2,0,0.1}
\definecolor{lightgray2}{cmyk}{0.1,0.1,0,0.1}

\makeatletter
\newlength{\apb@width}
\newcommand{\autoparbox}[2][c]{\settowidth{\apb@width}{#2}\parbox[#1]{\apb@width}{#2}}

\makeatother

\numberwithin{equation}{section}

\def\beq{\begin{equation}}
\def\eeq{\end{equation}}

\def\bea{\begin{eqnarray}}
\def\eea{\end{eqnarray}}

\def\Tr{{\rm Tr}}

\def\beq{\begin{equation}}
\def\eeq{\end{equation}}
\def\be{\begin{equation}}
\def\ee{\end{equation}}
\def\bea{\begin{eqnarray}}
\def\eea{\end{eqnarray}}

\def\Tr{{\rm Tr}}

\def\0{{\vec{0}}}
\def\p{{\vec{p}}}

\def\p{{\bf p}}

\DeclareRobustCommand{\SkipTocEntry}[4]{}

\def\t{\tau}
\def\p{\phi}

\def\beq{\begin{equation}}
\def\eeq{\end{equation}}

\def\ba#1\ea{\begin{align}#1\end{align}}
\def\bg#1\eg{\begin{gather}#1\end{gather}}
\newcommand{\bseq}{\begin{subequations}}
\newcommand{\eseq}{\end{subequations}}

\renewcommand{\t}{\Tilde}

\DeclareSymbolFont{extraup}{U}{zavm}{m}{n}
\DeclareMathSymbol{\varheart}{\mathalpha}{extraup}{86}
\DeclareMathSymbol{\vardiamond}{\mathalpha}{extraup}{87}

\def\({\left(}
\def\){\right)}
\def\[{\left[}
\def\]{\right]}

\begin{document}

\begin{titlepage}

\setcounter{page}{1} \baselineskip=15.5pt \thispagestyle{empty}

\vbox{\baselineskip14pt
%\hbox{hep-th/0000000}
}
{~~~~~~~~~~~~~~~~~~~~~~~~~~~~~~~~~~~~
~~~~~~~~~~~~~~~~~~~~~~~~~~~~~~~~~~
~~~~~~~~~~~ }

\bigskip\

\vspace{2cm}
\begin{center}
{\fontsize{19}{36}\selectfont  
{\sc De Sitter Holography and\\
\vspace{0.2cm}
Entanglement Entropy}
%SITP-18/02
}
\end{center}

\vspace{0.6cm}

\begin{center}
Xi Dong$^1$, Eva Silverstein$^2$, Gonzalo Torroba$^3$
\end{center}

%\vspace{0.2cm}

\begin{center}
\vskip 8pt

\textsl{
\emph{$^1$Department of Physics, University of California, Santa Barbara, CA 93106}}
\vskip 7pt
\textsl{\emph{$^2$Stanford Institute for Theoretical Physics, Stanford University, Stanford, CA 94306}}
\vskip 7pt
\textsl{ \emph{$^3$Centro At\'omico Bariloche and CONICET, Bariloche, Argentina}}

\end{center}

\vspace{0.5cm}
\hrule \vspace{0.1cm}
\vspace{0.2cm}
{ \noindent \textbf{Abstract}
\vspace{0.3cm}

We propose a new example of entanglement knitting spacetime together, satisfying a series of checks of the corresponding von Neumann and Renyi entropies.  The conjectured dual of de Sitter in $d+1$ dimensions involves two coupled CFT sectors constrained by residual $d$-dimensional gravity.  In the $d=2$ case, the gravitational constraints and the CFT spectrum are relatively tractable.  We identify a finite portion of each CFT Hilbert space relevant for de Sitter.  Its maximum energy level coincides with the transition to the universal Cardy behavior for theories with a large central charge and a sparse light spectrum, derived by Hartman, Keller, and Stoica.
Significant interactions between the two CFTs, derived previously for other reasons, suggest a maximally mixed state upon tracing out one of the two sectors; we derive this by determining the  holographic Renyi entropies.  The resulting entanglement entropy matches the Gibbons-Hawking formula for de Sitter entropy, including the numerical coefficient.  Finally, we interpret the Gibbons-Hawking horizon entropy in terms of the Ryu-Takayanagi entropy, and explore the time evolution of the entanglement entropy.

\vspace{0.4cm}

 \hrule

\vspace{0.6cm}}
\end{titlepage}

\tableofcontents

\section{Introduction}

\begin{figure}[htbp]
\begin{center}
\includegraphics[width=12cm]{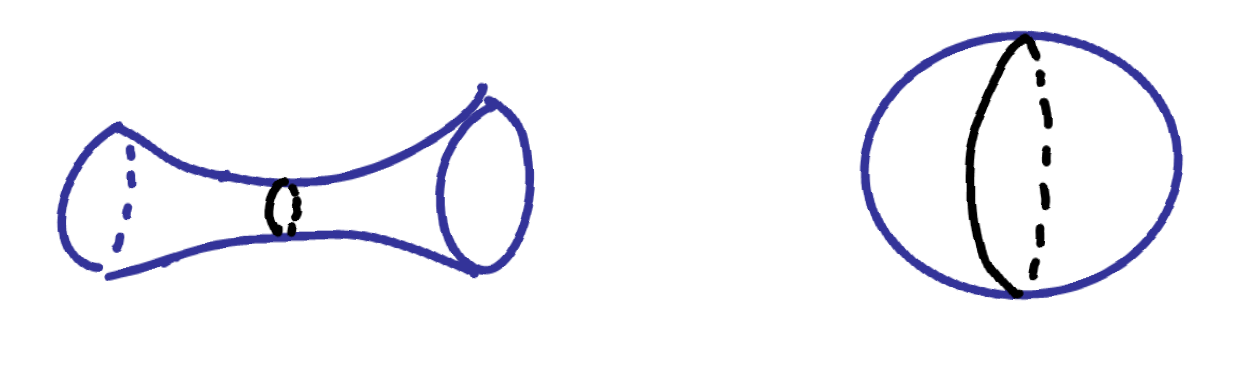}
\end{center}
\caption{On the left, we show an extremal surface (in black) bisecting a spatial slice (blue) of the eternal AdS black hole in $d=2$.  It corresponds to the entanglement entropy obtained by tracing over one of the CFTs in the thermofield double state (\ref{BHcase}).  On the right, we depict a spatial slice of $dS_3$ (\ref{dSslicing}) at the moment of time symmetry $\tau=0$, bisected by the extremal surface (in black) at the ultraviolet slice $w=\pi\ell_{dS}/2$.  Its area corresponds to the entanglement entropy obtained by tracing over one of the two coupled CFTs in the dual of de Sitter.  Given the maximally mixed density matrix derived from the Renyi entropies, we match the entanglement entropy, including its numerical coefficient. 
The entropy of the full circle is also the Gibbons-Hawking horizon entropy for an observer ${\cal O}$ momentarily at the deep IR end of one or the other warped throat, at $w=0$ or $\pi \ell_{dS}$ and $\tau=0$.  That observer has no interaction with the second CFT, and must trace over it.  See Fig.~\ref{Penrose} below for the corresponding space-time diagram.}
\label{volumeRT}
\end{figure}

The relations between quantum entanglement and spacetime geometry are an important part of the holographic dictionary.  This includes the Ryu-Takayanagi (RT) and Hubeny-Rangamani-Takayanagi (HRT) prescriptions \cite{RTHRT}\ which reproduce field theoretic entanglement entropy, including the area law dominated by high energy modes.\footnote{Generalizations beyond the GR limit appeared in \cite{Xietal}.}  An iconic example of an entangled state with an interesting gravitational description is the thermofield double state dual to the two-sided AdS black hole \cite{JuanTwoSided, VanR, EREPR},
\be\label{BHcase}
|\Psi\rangle = \sum_n e^{-\beta E_n/2} |n\rangle |n \rangle \,.
\ee 
Tracing out one of the two CFTs over the whole spatial volume of the CFT leads to an entanglement entropy that scales like the volume.  
On the gravity side, this is captured by the area of the extremal surface bisecting the two sides of the Einstein-Rosen bridge, as illustrated in the left panel of Fig.~\ref{volumeRT}.  
In this example, entanglement is in some sense dominated by the scale of the temperature, as higher energy components of the state (\ref{BHcase}) are Boltzmann suppressed.
The interpretation admits further checks and applications, some extending the wormhole \cite{ShenkerStanford}\ and others shortening it \cite{Traversable}, leading to new approaches to problems in many-body dynamics and black hole physics.  Other proposals for volume law entanglement have appeared, obtained by tracing over a range of energy scales \cite{MarkVradial}\ or over certain internal sectors \cite{sphereexample}. 

These examples involve a joining of different regions of space via entanglement  \cite{VanR}.  
In this note, we propose that de Sitter (dS) holography
realizes this general idea in a new way, admitting a counting of de Sitter entropy \cite{GibbonsHawking}\ which matches between the two sides of the duality.
In contrast to (\ref{BHcase}), the dominant CFT energy levels contributing to the entangled state are at the ultraviolet end of the relevant spectrum.  As we review shortly, the dual of de Sitter involves two lower-dimensional matter sectors that are cut off and coupled to each other, constrained by residual lower-dimensional gravity.
We identify the finite dimensional Hilbert space corresponding to each cutoff matter sector.  The Renyi entropies show that the reduced density matrix obtained by tracing out one of the sectors is maximally mixed (see the right panel of Fig.~\ref{volumeRT}).  This leads to a von Neumann entropy which matches between the two sides, including the numerical coefficient, and provides a dual description of the Gibbons-Hawking horizon entropy.  In what follows, we will initially focus on the $d=2$ case dual to $dS_3$, for which the analysis is technically simpler,\footnote{It is also in this dimensionality that the most explicit string theoretic de Sitter construction was formulated, uplifting the D1-D5 AdS/CFT duality \cite{Micromanaging}.} and we will then generalize to higher dimensions.  We will also explore the time evolution and late time behavior of the entanglement entropy between the two sectors.

\section{Setup}\label{sec:setup}

There are several mutually consistent proposals for cosmological holography \cite{dSCFT, Hologravity, Micromanaging, FRWFRW, FRWCFT, TASIcosmo, musings}.  The dS/dS duality starts from a simple observation \cite{Hologravity}: 
$dS_{d+1}$ is a warped compactification down to a $d$-dimensional theory on $dS_{d}$:
\bea\label{dSslicing}
ds^2_{dS_{d+1}} &=& dw^2+\sin^2\(\frac{w}{\ell_{dS}}\) ds^2_{dS_d} \nonumber\\
&=& dw^2+\sin^2\(\frac{w}{\ell_{dS}}\) \left[-d\tau^2 + \ell_{dS}^2 \cosh^2\frac{\tau}{\ell_{dS}} d\Omega_{d-1}^2\right]\,.
\eea
This exhibits two highly redshifted regions at $w\to 0$ and $w\to \pi \ell_{dS}$, which are each equivalent to the gravity dual of the low energy regime of a CFT on $dS_{d}$.  This is in contrast with the AdS/CFT duality in $dS_d$ slicing, for which the warp factor is $\sinh^2(w/\ell_{AdS})$ instead of $\sin^2(w/\ell_{dS})$:
\be\label{AdSdS}
ds^2_{AdS_{d+1}}= dw^2+\sinh^2\(\frac{w}{\ell_{AdS}}\) \left[-d\tau^2 + \ell_{AdS}^2 \cosh^2\frac{\tau}{\ell_{AdS}} d\Omega_{d-1}^2\right]\,.
\ee
On the other hand, in de Sitter the most UV scale is finite, occurring at $w_{uv}=\pi\ell_{dS}/2$, with the warp factor
\be\label{UVwarp}
\sqrt{g_{00}}|_{uv}=\sin \frac{w_{uv}}{\ell_{dS}}=1.
\ee
This indicates a semi-holographic dual description as two matter sectors cut off at the energy scale corresponding to (\ref{UVwarp}) and coupled to each other as well as to a residual $d$-dimensional gravity.\footnote{$dS_{d+1}$ is a very special Randall-Sundrum \cite{RS}
system, with no explicit Planck brane and a highly constrained holographic RG flow \cite{HoloRG}.
The $d$-dimensional gravity is dynamical at finite times, and we will be concerned with the period
over which the spacetime is classically de Sitter.  However, it is worth noting that
after decaying to a runaway toward zero cosmological constant, the dual gravity ultimately decouples \cite{FRWFRW}.} 

The same statement arises independently from the basic structure of metastable de Sitter solutions in string theory \cite{Micromanaging, TASIcosmo}.  The presence of two isomorphic matter sectors follows from the uplift of the AdS/CFT brane construction, with the metastability of string solutions entering into this in an essential way \cite{Micromanaging, TASIcosmo}.  The proposed dS/CFT
duality \cite{dSCFT}\ also ultimately consists of 2 identical matter sectors coupled to $d$-dimensional gravity \cite{HarlowStanford}.  This consistent picture seems unlikely to be pure coincidence, and we will pursue its consequences further in the present work.
  
There is a suite of calculations checking the basic structure of the duality \cite{Hologravity, Micromanaging, FRWFRW, HoloRG}.
Of course, the tools are more limited in this cosmological context.  One aspect of this is that the dual matter sectors only behave as a pair of CFTs at sufficiently low energies; they are deformed by irrelevant operators.\footnote{This includes the $T\bar T$ deformation \cite{ZTTbar}, but is not limited to that.  It is interesting to note that gravitational dressing in two dimensions appears to reproduce the effect of  the $T\bar T$ deformation analyzed using non-gravitational methods in \cite{ZTTbar}\ (see for example \cite{TTbarJT, AkiBill, Cardy}).
As noted in \cite{MarolfKraus}, the holographic dual of the $T\bar T$ deformation \cite{Verlinde} requires further information about the interactions than is captured purely by \cite{ZTTbar}.  In our case this is reflected in part in the couplings (\ref{Smix}).}

This will actually play an important role here.
The transmission coefficient between the two low energy throats \cite{Hologravity}\ reveals that the two theories are coupled via irrelevant interactions of the form 
\be\label{Smix}
S_{mix}\sim \int d^d x\sqrt{-g} \, \lambda \, \ell_{dS}^{2\Delta-d}{\cal O}_1{\cal O}_2 +\dots
\ee
where $g_{\alpha\beta}$ is the $d$ dimensional metric, classically $dS_d$.  In Appendix \ref{app:mix} below, we will study $\lambda$ at the Gaussian level for large-$N$ factorized fields.  The mass scale suppressing these irrelevant operators is of order the Hubble scale $1/\ell_{dS}$ of the de Sitter spacetime.\footnote{Similar statements apply to more general 2-throated Randall-Sundrum systems, where such calculations were originally performed \cite{Tunneling}.}  The `$\dots$' in (\ref{Smix}) indicates non-Gaussian effects, which become important at the energy levels $\Delta\sim c$ which will be of interest here.  The strong interactions between the two matter sectors plausibly generate a long-lived state which is nearly maximally mixed, a feature that we will probe directly below using Renyi entropies.\footnote{Previous examples of interactions leading to strongly mixed states include \cite{sphereexample}\ and references therein; we provide a toy model in the Appendix \ref{app:spin}.}

%%%%%%%%%%%%%%%%%%%%%%%%%%%%%%%%%%%%%
%%%%%%%%%%%%%%%%%%%%%%%%%%%%%%%%%%%%%

\section{CFT state space and entropy in the 2d theory}\label{sec:2d}

\subsection{CFT with a cutoff}

For $d=2$, the two-dimensional dual theory contains two CFTs, each with a large central charge $c$, and irrelevant interactions including mixings (\ref{Smix}), all coupled to $d=2$ gravity.  Our first question is what sector of the Hilbert space of the CFTs comes in.

Recent work has constrained the density of states in two-dimensional CFTs with a large central charge and a sparse light spectrum \cite{CFTspectrum}, such as those dual to large-radius gravity.  For simplicity, we focus on spinless states; the generalization is interesting but does not affect our basic results.  

There is a transition at CFT energies
\be\label{Deltatrans}
E_{CFT *} = \Delta -\frac{c}{12} = \frac{c}{12}, ~~~~~ \Delta = \frac{c}{6}
\ee
at which the entropy begins to follow the Cardy formula, given a sparse light spectrum \cite{CFTspectrum}.  
This is exactly the relevant part of the Hilbert space for $dS$.

To see this, start with the holographic dual of one of the CFTs living on global $dS_2$.  In full, this is given by the patch of $AdS_3$ covered by (\ref{AdSdS}).
The CFT states (\ref{Deltatrans}) correspond to the spinless BTZ black hole states in global $AdS_3$ with horizon radius 
\be\label{rhell}
r_{h*}=\ell_{AdS}.  
\ee
They are at the threshold where black holes begin to dominate the thermodynamics.  What is important about them for our present purposes is that their horizon intersects the $\tau=0$ neck of the $dS_2$ slices in (\ref{AdSdS}) at unit warp factor, $\sqrt{g_{00}}=\sinh(w/\ell_{AdS})=1$.
In the $dS_3/dS_2$ system (\ref{dSslicing}), the most ultraviolet scale is at unit warp factor, $\sin(w/\ell_{AdS})=1$, as pointed out in (\ref{UVwarp}).  This suggests that we should restrict the Hilbert space of each of our two CFTs to $E_{CFT}\le E_{CFT *}$.

This amounts to a sharp cutoff on the CFT at a finite radial scale at the level of each matter sector, before mixing them or coupling to $2d$ gravity.  We will see that this produces the correct entropy, given the structure of the density matrix derived below, providing concrete evidence for this simple cutoff prescription.  However, we wish to note an ambiguity here:  there may be irrelevant operators deforming each CFT, independently of their mixing (\ref{Smix}), which could lead to a deformation of the dual geometry below the radial cutoff at the level of each individual warped throat.  This would generically not produce the same sharp agreement, but there may be special flows that also fit the facts.  In \cite{HoloRG}, we found that if one treats one side, e.g. $w<w_{uv}$ in $dS_{d+1}$, as the gravity dual of a matter theory, this theory has a highly constrained holographic Wilsonian RG flow:  single-trace operators do not flow, and higher traces are determined in terms of lower ones.  There is also an upper bound on the mass allowed in such a throat, which coincides with the bound we were led to above.\footnote{In higher dimensions the bound corresponds to the limit where the black hole and cosmological horizons coincide -- see e.g.~\cite{musings} for a review of this limit.}  

The number of states up to the bound (\ref{Deltatrans}) is given by
\be\label{Strans}
{\rm dim} (H_{\Delta \le c/6}) = e^{\pi c/3}
\ee
to the leading order in the large-$c$ limit.  This can be obtained from black hole states in the microcanonical ensemble near the bound $E_{CFT *}$.
Below in \S\ref{sec:higher}, we will generalize this to higher dimensions, finding that the energy levels go up to the Hawking-Page transition in the canonical ensemble, with the number of states given by (\ref{SBH}), generalizing (\ref{Strans}).

Having identified the part of the CFT Hilbert space that comes into the holographic description of a long-lived de Sitter geometry, we should stress that the full system has more degrees of freedom which enter in the later decay of de Sitter \cite{FRWFRW,HoloRG}, in contrast to proposals with a finite total number of degrees of freedom such as \cite{Nbound}.  In both cases, however, the finite number of states relevant for the dual description of a long-lived de Sitter configuration is related to backreaction on its geometry. 

In Appendix \ref{app:liouville}, we spell out the role of the Liouville gravity and how it leads to the $dS_2$ evolution.

\subsection{Structure of the density matrix}

The full system contains mixing interactions between the two CFTs, such as (\ref{Smix}).
Let us denote the ground state wavefunction of the full system by $|\Psi\rangle$, and the two cut-off CFT sectors by QFT$_{1,2}$. 
The density matrix for the first sector is given by
\be\label{eq:densitymatrix}
\rho_1= \text{Tr}_{\text{QFT}_2}(|\Psi\rangle \langle \Psi|)\,.
\ee
We will be interested in the properties of $\rho_1$, specifically its von Neumann entropy (or entanglement entropy)
\be\label{SVN}
{\cal S}= - \text{Tr}( \rho_1 \log \rho_1)
\ee
and its Renyi entropies 
\be\label{Renyidef}
{\cal S}_n = \frac{1}{1-n}\log \text{Tr} \,\rho_1^n\,,
\ee
which capture the amount of entanglement between the two matter sectors.

We might expect that the strong direct interactions between the two sectors leads to a long-lived state of the full system that is approximately maximally entangled, giving a maximally mixed state for the first sector:
\be\label{rhomixeddef}
\rho_{1, max} \simeq {\mathbb{I}}\frac{1}{{\rm dim} (H_{\Delta \le c/6})}
\ee
where $\mathbb{I}$ denotes the identity matrix.  This maximal mixing is not something that we are able to compute directly in the lower-dimensional dual theory, but there are toy models that exhibit such a mixing effect of interactions.\footnote{We illustrate this with a spin lattice system in Appendix~\ref{app:spin}.}
If this is the case, the entanglement entropy (\ref{SVN}) obtained by tracing out the second QFT, and the corresponding Renyi entropies (\ref{Renyidef}), are all given by 
\be\label{Smatch}
\rho_1\simeq \rho_{1, max}\quad
\Leftrightarrow\quad
{\cal S}={\cal S}_n= \log \dim( H_{\Delta\le c/6})=\frac{\pi c}{3}\,,
\ee
where we used the finite dimension of the relevant Hilbert space (\ref{Strans}).

We will show in \S\ref{sec:gcal} that the $(d+1)$-dimensional gravity side produces exactly the behavior in (\ref{Smatch}).  
In particular, the full set of Renyi entropies computed there matches (\ref{Smatch}) and implies maximal mixing, up to $1/c$ corrections.

%%%%%%%%%%%%%%%%%%%%%%%%%%%%%%%%%%%%%%%%
%%%%%%%%%%%%%%%%%%%%%%%%%%%%%%%%%%%%%%%%
\section{Holographic entanglement and Renyi entropies in $dS_{d+1}/dS_d$}\label{sec:GRent}

\subsection{Gravity side calculation}\label{sec:gcal}

In a holographic theory, the entanglement entropy for any boundary region is given by the area of an appropriate codimension-2 extremal surface in the bulk \cite{RTHRT}:
\be
{\cal S} = \frac{A_{\text{ext}}}{4G_{d+1}}.
\ee
In our highly symmetric $dS_{d+1}$ spacetime (\ref{dSslicing}), there is a codimension-2 extremal surface at $w=\pi \ell_{dS}/2$ and at the moment of time reflection symmetry $\tau=0$ in the global $dS_d$. This gives
\be\label{VolEnt}
{\cal S} = \frac{V_{d-1}}{4 G_{d+1}}\,,\qquad
V_{d-1} = \frac{2\pi^{d/2} \ell_{dS}^{d-1}}{\Gamma(d/2)}
\ee
where $V_{d-1}$ is the area (volume) of the extremal surface at the $S^{d-1}$ neck.  Here the entangling region is the whole spatial volume of one of the two identical matter sectors, and (\ref{VolEnt}) corresponds to tracing out completely the other matter sector, giving the von Neumann entropy (\ref{SVN}) of the density matrix (\ref{eq:densitymatrix}).  

The extremal surface does not drop into either side of the bulk, but rather stays on the UV slice as depicted in Fig.~\ref{volumeRT}, leading to entanglement over the whole volume $V_{d-1}$ of the $S^{d-1}$ neck.  This is a consequence of the fact that the two throats are joined smoothly, with a warp factor $\sqrt{g_{00}} = \sin(w/\ell_{dS})$ whose first derivative vanishes at the UV slice.  We discuss this calculation in more detail in  Appendix~\ref{app:EE}.  Similar results were found in \cite{Narayan:2017xca}.

In the $d=2$ case discussed in \S\ref{sec:2d}, we can express (\ref{VolEnt}) in terms of the dual variables:
\be\label{Stwo}
{\cal S} = \frac{2\pi \ell_{dS}}{4 G_3} = \frac{\pi c}{3}\,,
\ee
where we have used the fact that the proper length of the neck is $2\pi \ell_{dS}$ as well as the relation
\be\label{crel}
c=\frac{3\ell_{AdS}}{2 G_3}
\ee
from the $AdS_3/CFT_2$ dictionary.  We have identified the curvature radii:
\be\label{ells}
\ell_{dS}=\ell_{AdS}\,.
\ee 
This is because each matter sector reverts in the infrared to a CFT on $dS_2$, with a small warp factor $\sin^2(w/\ell_{dS})\ll 1$.  This warp factor in $dS_3$ (\ref{dSslicing}) is indistinguishable from a small warp factor $\sinh^2(w/\ell_{AdS})$ in $AdS_3$ (\ref{AdSdS}), once we identify the two curvature radii as in (\ref{ells}).  

So far, the gravity side reproduces (\ref{Smatch}) precisely for the von Neumann entropy.
We can calculate the Renyi entropies (\ref{Renyidef}) 
which contain further information about the full spectrum of the density matrix.  Following~\cite{Lewkowycz:2013nqa,Dong:2016fnf}, we will work with a natural generalization of the Renyi entropies
\be\label{Srentilde}
\widetilde {\cal S}_n \equiv n^2\partial_n \left(\frac{n-1}{n}{\cal S}_n\right) = -n^2\partial_n\left(\frac{1}{n}\log \Tr\, \rho_1^n \right)
\ee
which has a simple gravity dual in holographic theories and is given by
\be
\widetilde{\cal S}_n = \frac{A(C_n)}{4G_{d+1}}
\ee
where $A(C_n)$ is the area of a codimension-2 cosmic brane $C_n$ with tension
\be\label{branetension}
T_n = \frac{n-1}{4n G_{d+1}}
\ee
in the Euclidean bulk theory.
Note that in the limit $n \to 1$, the generalized Renyi entropy $\widetilde {\cal S}_n$ approaches the entanglement entropy, while the cosmic brane becomes tensionless and does not backreact on the bulk geometry.  It becomes a probe brane and settles at the location of the extremal surface.  In contrast, for generalized Renyi entropies with $n>1$ one needs to include the backreaction of the cosmic brane on the geometry, creating a conical deficit angle
\be\label{defangle}
\Delta\phi = 2\pi \frac{n-1}{n}\,.
\ee

In our de Sitter theory, this turns out to be a simple calculation.  Again, by symmetry the cosmic brane wraps the $S^{d-1}$ neck between the two warped throats.  Let us consider the $dS_3/dS_2$ case first, for which the Euclidean geometry is a 3-sphere.  We can describe it in terms of complex variables $z_1, z_2$ as
\be\label{Sthree}
|z_1|^2+|z_2|^2=\ell_{dS}^2.
\ee
The cosmic brane with its deficit angle (\ref{defangle}) arises at the fixed locus of the orbifold
\be\label{orb}
(z_1, z_2) \simeq (e^{2\pi i/n} z_1, z_2)\,.
\ee
The fixed locus is the circle 
\be\label{fixedcircle}
z_1=0, ~~~~ |z_2|^2=\ell_{dS}^2\,.
\ee
The important feature for us here is that the fixed locus has the same length $2\pi \ell_{dS}$ for all $n$.  The same feature holds in general dimensions by a similar argument.

From this we find that the generalized Renyi entropies are independent of $n$: $\widetilde{\cal S}_n = {\cal S}$.  Integrating (\ref{Srentilde}) over $n$, we find the conventional Renyi entropies ${\cal S}_n$ are the same:
\be\label{sns}
{\cal S}_n = {\cal S} = \frac{\ell_{dS}^{d-1}}{4 G_{d+1}} \frac{2\pi^{d/2}}{\Gamma(d/2)}\,.
\ee
This implies a maximally mixed density matrix $\rho_1$ (to the leading order in the large-$c$ limit), with 
Renyi entropies that precisely reproduce (\ref{Smatch}).\footnote{Similar maximal mixing in de Sitter was found in \cite{Nomura:2017fyh} although the interpretation was different from ours.}.

\subsection{Gibbons-Hawking entropy}

\begin{figure}[htbp]
\begin{center}
\includegraphics[width=8cm]{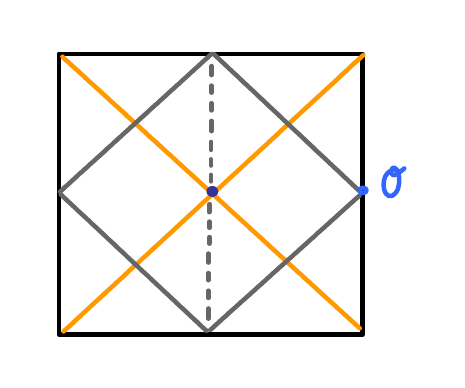}
\end{center}
\caption{The Penrose diagram of $dS_{d+1}$, with the UV slice $w=\pi \ell_{dS}/2$ of the geometry indicated by the dashed grey line.  The solid grey lines indicate the two deep IR regions, $w=0$ and $w=\pi\ell_{dS}$.  The RT surface $S^{d-1}$ is in the center of the diagram, indicated as a dark blue point. 
Its area determines the entanglement entropy between the two matter sectors. and tracing out the second sector gives a maximally mixed density matrix $\rho_1$ for the first.  It is also the Gibbons-Hawking horizon entropy for an observer momentarily at the deep IR end of the second warped throat at $\tau=0$, indicated in light blue.  That observer has no interaction with the first matter sector and must trace it out.}
\label{Penrose}
\end{figure}

The entropy (\ref{VolEnt}) is equal to the horizon entropy for an observer momentarily at the deep infrared end of one of the two warped throats (i.e. at the north or south pole of the $dS_{d+1}$ spacetime):  
\be\label{SdS}
{\cal S} = \frac{V_{d-1}}{4 G_{d+1}} = {\cal S}_\text{Gibbons-Hawking}\,.
\ee
Such an observer must trace out the matter sector describing the other warped throat, as shown in Fig.~\ref{Penrose}.. This gives us an interpretation of the Gibbons-Hawking entropy, including its numerical coefficient.

\subsection{Volume law}\label{sec:volumecomment}

In our calculations above, we have worked on the full volume of one of the two matter sectors, tracing out the other sector.  
The RT calculation and generalizations to the Renyi entropies can also be done formally on the $(d+1)$-dimensional gravity side for subregions of size smaller than $\ell_{dS}$, which do not cover the whole volume $V_{d-1}$. As shown in Appendix~\ref{app:EE}, the result is again a volume law (\ref{eq:volume}). 

This volume law for subsystems of one of the two matter sectors (say QFT$_1$) is fully consistent with a maximally mixed density matrix $\rho_1$ for QFT$_1$.  On the other hand, it is not immediately clear how to identity the subsystem corresponding to a subregion in QFT$_1$ at the level of the Hilbert space, due to the imposition of the energy cutoff (\ref{Strans}).
The AdS/CFT and dS/dS dualities contain $\sim c$ degrees of freedom in a region of size $\ell_{(A)dS}$, as explained in \cite{SusskindWitten}.  One can work with bulk fields that interact weakly, but these do not dominate the entropy.  As we saw above in $d=2$, 
the entanglement entropy is dominated by states with $\Delta \simeq c/6$ in the two CFTs, which correspond to black holes with horizon radius $\ell_{AdS}$ on the gravity side. Nonetheless, it would be interesting to understand the subsystems corresponding to smaller regions.  It may hold clues about locality at scales below the de Sitter curvature radius.

\section{Late times and HRT}\label{sec:late}

By going to later times in the global $dS_d$ space we can incorporate more Hubble patches of proper size $\ell_{dS}$.  Points separated by a fixed coordinate distance eventually lose causal contact, and we may expect a volume law for the entropy, of order $c$ per Hubble patch.  In this section, we will see how this behavior comes out, by calculating the entanglement entropy of a region at a general time in the dS/dS duality.  We will restrict our attention to the long time period over which a metastable de Sitter configuration has not yet decayed; it would be interesting to generalize our discussion to the case of \cite{FRWFRW} in the future.

Let us illustrate this in the $dS_3/dS_2$ case.  We set $\ell_{dS}$ to 1 in this section for simplicity.
Consider when the entangling region is an interval of size $\phi$ on the UV slice $w=\pi/2$ at a general time $\tau_0$.  The holographic entanglement entropy of this region is given by the area of an extremal surface (a geodesic in our $d=2$ case) homologous to the region, according to the covariant HRT prescription \cite{RTHRT}.  Let us assume $\tau_0\geq 0$ without loss of generality.

\begin{figure}[htbp]
\centering
\includegraphics[width=0.5\columnwidth]{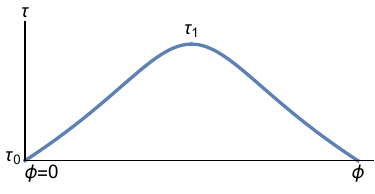}
\caption{HRT extremal surface for an interval of size $\p$ at a general time $\tau_0$.}
\label{fighrt}
\end{figure}

As in our other calculations for holographic entropies, the symmetry and smoothness at the UV slice implies that the extremal surface (geodesic) lies on the UV slice at $w=\pi/2$.  The geodesic is moreover symmetric under a reflection exchanging the two end points of the interval.  The fixed point of this reflection defines a turning point $\tau_1$ where the $\tau$ coordinate reaches an extremum.  Parameterizing (one half of) the geodesic by $\p(\tau)$, we find its area (length):
\be\label{afn}
A = 2\int_{\tau_0}^{\tau_1} d\tau \sqrt{\dot\p^2 \cosh^2\tau-1}.
\ee
As a result of the rotational symmetry in the $\phi$ direction, extremizing (\ref{afn}) gives
\be
\frac{\dot\p \cosh^2\tau}{\sqrt{\dot\phi^2 \cosh^2\tau-1}} = C
\ee
where $C$ is a constant.  The $\tau$ coordinate reaches an extremum at the turning point $\tau_1$, giving $\dot\p(\tau_1)=\infty$ and leading to
\be
C = \cosh\tau_1.
\ee
From this we find
\be\label{phidot}
\dot\p = \frac{\cosh\tau_1}{\cosh\tau \sqrt{\cosh^2\tau_1 - \cosh^2\tau}}.
\ee
This means $\tau_1\geq \tau_0$ and that the geodesic bends towards later times, rather than earlier times, as illustrated in Fig.~\ref{fighrt}.
Integrating (\ref{phidot}) from $\tau_0$ to $\tau_1$ and setting it to $\p/2$, we find
\be\label{patan}
\p = 2\arctan\(\frac{\sqrt{\cosh^2\tau_1 - \cosh^2\tau_0}}{\sinh\tau_0 \cosh\tau_1}\).
\ee
We use the convention where the range of $\arctan$ is $[0,\pi]$ instead of $[-\pi/2,\pi/2]$.  The above equation determines $\tau_1$ from $\tau_0$ and $\p$.

The extremal area (length) is
\be\label{aatan}
A = 2\arctan\(\frac{\sqrt{\cosh^2\tau_1 - \cosh^2\tau_0}}{\sinh\tau_0}\) = 2\arctan\(\tan\frac{\p}{2}\cosh\tau_1\).
\ee
From this we determine the entanglement entropy ${\cal S}=A/(4G_3)$.  There is a subtlety in taking the late time limit, which contains some interesting physics related to the causal structure of de Sitter.

To see the subtlety, we solve (\ref{patan}) and find $\tau_1$ in terms of $\tau_0$ and $\p$:
\be\label{ctos}
\cosh\tau_1 = \frac{\cosh\tau_0}{\sqrt{1-\sinh^2\tau_0 \tan^2(\p/2)}}.
\ee
This blows up as
\be
\sinh\tau_0 \to \cot \frac{\p}{2},
\ee
or
\be
\tau_0 \to \tau_*(\p) \equiv \text{arcsinh}\(\cot\frac{\p}{2}\).
\ee
What happens in this limit is that the turning point $\tau_1$ reaches the future infinity.\footnote{As mentioned above, the de Sitter description only persists until a very large but finite $\tau_1$, given that metastable de Sitter solutions eventually decay.  It would be interesting to see the effect of this on the extremal surface.}  From (\ref{phidot}), we see that the extremal surface is lightlike except at the turning point in this limit.

If we pushed our equations beyond this limit, we would be prescribing an entangling region that is not in causal contact.
Mathematically, the corresponding extremal surface would become complex\footnote{This question was also raised in \cite{dSCFTcomplex}, which considers holography in finite regions as well.} as can be seen from (\ref{ctos}), similar to situations encountered in other contexts \cite{ComplexHRT}.

\begin{figure}[htbp]
\begin{center}
\includegraphics[width=15cm]{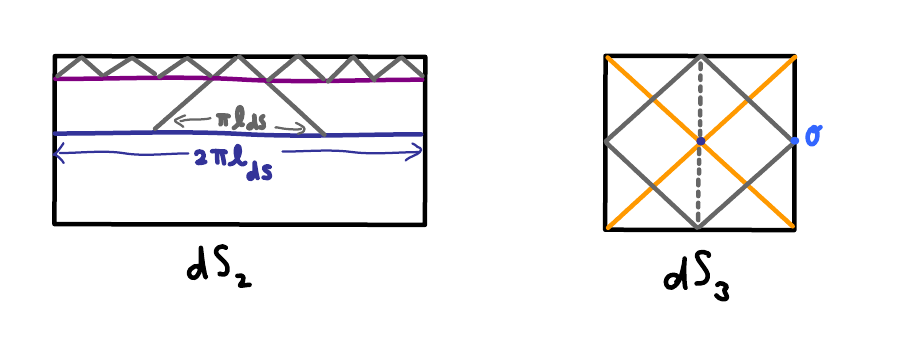}
\end{center}
\caption{The left panel contains the $dS_2$ Penrose diagram; the vertical sides are identified.  The dark blue line at $\tau=0$ corresponds to the central dot (a circle of radius $\ell_{dS}$) in the $dS_3$ diagram from Fig. \ref{Penrose}, reproduced here on the right.  This full circle of size $2\pi\ell_{dS}$ is the horizon for the observer ${\cal O}$ on the far right, causally connected to it through the $dS_3$ bulk.  Within the global $dS_2$ geometry, the segments with extremal area $\pi\ell_{dS}$ between the end points are the largest for which the endpoints can communicate, as derived in the text.  On a late-time slice, indicated in purple, we show a string of causal segments whose extremal area is $\pi\ell_{dS}$ and entanglement entropy is $\pi c/6$.}
\label{Causal}
\end{figure}

Instead, we may string together separate causally connected segments, as depicted in Fig.~\ref{Causal}.
For each such segment, restoring the de Sitter radius we find the extremal area (length) is $\pi\ell_{dS}$ as given by the $\tau_1\to\infty$ limit of (\ref{aatan}), and the entanglement entropy per causal segment is
\be\label{shalf}
{\cal S}_{\pi\ell_{dS}}= \frac{\pi \ell_{dS}}{4G_3} = \frac{\pi c}{6}.
\ee
Note that this agrees with the entropy that we have calculated using the RT surface at the neck in (\ref{Stwo}): there we integrated over the full $2\pi\ell_{dS}$, obtaining $\pi c/3$.  Here we get the same result by summing the entropies of the two causal segments at $\tau=0$.

%%%%%%%%%%%%%%%%%%%%%%%%%%%%%%%%%%%%%%%%%%%%%%%%%%
%%%%%%%%%%%%%%%%%%%%%%%%%%%%%%%%%%%%%%%%%%%%%%%%%%
%%%%%%%%%%%%%%%%%%%%%%%%%%%%%%%%%%%%%%%%%%%%%%%%%%
\section{Higher dimensions}\label{sec:higher}

We can extend the entropy matching to higher dimensions by generalizing (\ref{rhell}), although there is no independent CFT calculation of this transition as in $d=2$ \cite{CFTspectrum}.  In short, the Hawking-Page transition occurs universally at $r_{+*}=\ell_{AdS}$ in the canonical ensemble \cite{WittenHP}.  The corresponding energy scale in $(A)dS_{d+1}/dS_d$ is at a warp factor ${\rm sin(h)}(w/\ell_{AdS})=1$.  The logarithm of the number of states below this energy level in the CFT is given by the entropy of this black hole, 
\be\label{SBH}
\log({{\rm dim} H})={\cal S}= \frac{\ell_{AdS}^{d-1}}{4 G_{d+1}} \frac{2\pi^{d/2}}{\Gamma(d/2)} \,.
\ee
Given the well-established AdS/CFT correspondence, we can incorporate this as a statement about the CFT.
Therefore, a maximally entangled state of the two cut-off CFT sectors in the dS/dS duality leads to the same entanglement entropy and Renyi entropies as in (\ref{SBH}).  This agrees precisely with the gravitational prediction (\ref{sns}) given (\ref{ells}).

%%%%%%%%%%%%%%%%%%%%%%%%%%%%%%%%%%%%%%%%%%%%%%%%%%
%%%%%%%%%%%%%%%%%%%%%%%%%%%%%%%%%%%%%%%%%%%%%%%%%%
%%%%%%%%%%%%%%%%%%%%%%%%%%%%%%%%%%%%%%%%%%%%%%%%%%
\section{Summary and recap of the logic}\label{sec:recap}

Altogether, we have obtained a set of nontrivial matches among the Gibbons-Hawking, von Neumann, and Renyi entropies (with numerical coefficients), the structure of the density matrix, and the finite Hilbert space arising in the dS/dS correspondence.

It is worth recapping the logic and interpretation.
We have obtained the finite-dimensional Hilbert space relevant for the dS/dS duality in (\ref{Strans}) and (\ref{SBH}).  We can proceed from it in two ways, running the duality in either direction:

$\bullet$  Starting from the $d$-dimensional theory, we can hypothesize the maximal mixing of $\rho_1$ based on the existence of strong interactions between the two matter sectors.  That combined with (\ref{Strans}) and (\ref{SBH}) leads to the von Neumann entropy ${\cal S}$ and Renyi entropies ${\cal S}_n$ that match with the calculation on the gravity side, including the numerical coefficients.

$\bullet$  Alternatively,  we can use the $(d+1)$-dimensional gravity description to calculate the entanglement and Renyi entropies.  That implies the maximal mixing of $\rho_1$.  From (\ref{Strans}) and (\ref{SBH}), we match the value of the entropies, including the numerical coefficients.

The von Neumann and Renyi entropies of $\rho_1$ arise from tracing out the second matter sector, QFT$_2$.
This provides a new example of strong entanglement resulting in the joining of two holographic spacetimes.

The value of the entropies is the same as the Gibbons-Hawking entropy of the de Sitter horizon.  This fits in precisely as well: an observer who must trace out QFT$_2$ is situated at the far infrared end of the first warped throat.

\section{Discussion and future directions}\label{sec:concl}

In this work we studied the relation between spacetime geometry and entanglement of the holographic dual degrees of freedom, in de Sitter space.   We obtained a finite Hilbert space for each of the two identical matter sectors.  In $d=2$, this is specifically $\Delta\le c/6$, the level at which the Cardy formula kicks in for theories with sparse light spectra, as derived by traditional CFT methods in \cite{CFTspectrum}.   We combined this with the maximal mixing of the density matrix $\rho_1$ obtained by tracing out one of the two sectors, which was implied by the equality of the Renyi entropies that we calculated (and was suggested independently by the significant interactions between the matter sectors).   This produces the correct de Sitter entropy, including its numerical coefficient, with a simple interpretation as the entanglement entropy between the two identical matter sectors in the dual theory.

This provides strong evidence for the conjecture that de Sitter spacetime gives a new example of entanglement joining spacetime  \cite{JuanTwoSided,VanR,EREPR,MarkVradial,sphereexample}.
It is interesting to compare and contrast this with the  thermofield double/black hole case (\ref{BHcase}) \cite{JuanTwoSided}.  
Both involve entanglement over the whole spatial volume of two CFTs.  
In our case, the CFT is cut off at a finite scale, and entanglement occurs at the highest of this finite set of relevant energy levels and not at the scale of the temperature as in the thermofield double (\ref{BHcase}). This translates into the joining of the two subsystems at their most UV slice, and the Renyi entropies imply a maximally mixed density matrix $\rho_1$ for one of the subsystems.
A significant distinction from the eternal AdS black hole case is that here the two subsystems have large interactions. It is these interactions that lead to the maximal entanglement of the long-lived state corresponding to unexcited $dS_{d+1}$, similar to what happens in simple models such as the lattice model in Appendix \ref{app:spin}.
In the AdS black hole case, one can introduce interactions that lead in a novel traversability of the bridge \cite{Traversable}.   In the de Sitter case, the transmission between the two warped throats is an unsurprising consequence of the interactions \cite{Tunneling,Hologravity}.  

A crucial part of our analysis is the identification of the correct Hilbert space, carried out in \S\ref{sec:2d} and \S\ref{sec:higher}. In $d=2$, it is restricted to states with $\Delta \le c/6$ with the bound corresponding to black holes with horizon radius $r_h=\ell_{AdS}$ in the gravity dual of each CFT. These dominate the entropy, and cannot be interpreted in terms of weakly-interacting local fields. It is interesting to consider the familiar question of characterizing what happens at scales below the (A)dS curvature radius in the present context.  In Appendix \ref{app:EE}, we consider the entanglement entropy for regions with size smaller than $\ell_{dS}$ and still found a volume law. It is also possible to obtain a more refined estimate of the correlations between small subsystems by computing the mutual information. This calculation is done in Appendix \ref{app:EE}, with the result that the mutual information always vanishes to the leading order in $1/c$. This should be contrasted with the AdS/CFT case, where for regions that are sufficiently close to each other there is a Hagedorn-like transition and the mutual information becomes nonvanishing~\cite{Headrick:2010zt}. 

The Renyi entropies $\mathcal S_n$ for one of the two matter sectors in the dS/dS duality are independent of $n$ and equal to the von Neumann entropy.  This is in contrast with the case of a boundary region in AdS/CFT, where $\mathcal S_n$ typically depends nontrivially on $n$.  However, the equality of Renyi entropies holds in interesting toy models of holography based on tensor networks \cite{Pastawski:2015qua,Hayden:2016cfa} that exhibit the quantum error-correcting properties found in AdS/CFT \cite{Almheiri:2014lwa}, and in a certain sense the equality of Renyi entropies ``almost'' holds in AdS/CFT as well \cite{DHM}.  It would be very interesting to connect these special cases and further understand the structure of entanglement in gravitational states.

The cosmological setting opens up many new directions to understand the interplay between entanglement and gravity. It will be interesting to generalize the analysis to more general FRW cosmologies \cite{FRWFRW}, including the decays of de Sitter, using the HRT prescription.  
Here gravity decouples at late times and the number of degrees of freedom increases;  the Bousso entropy bound \cite{BoussoBound}\ asymptotically goes to infinity, along with the $d$-dimensional Planck mass.  
In Appendix \ref{app:liouville}, we incorporate the main role of gravitational dressing in producing a Wheeler-DeWitt solution describing $dS_2$.  It would be interesting to pursue its relation to the $T\bar T$ deformation \cite{ZTTbar, TTbarJT}.

Although we stress that the dS/dS correspondence and the arguments in this paper apply for a large radius de Sitter space \cite{Micromanaging}, it might be interesting to analyze the holographic dual in cases involving weakly interacting fields in the $d$-dimensional dual theory.  In those cases, one might be able to directly compute additional properties of the wavefunction yielding the maximally mixed density matrix.  This would need to be compared to an appropriate generalization of the RT prescription to gravity duals with a small curvature radius.

\bigskip

\noindent{\bf Acknowledgements}

\noindent We are grateful for discussions with Sergei Dubovsky, Victor Gorbenko, and Shamit Kachru on recent developments on 2d CFTs and gravity, and with Steve Shenker on the holographic framework.   We would also like to thank N. Bao, J. R. Bond, H. Casini, W. Cottrell, D. Marolf, M. Mirbabayi, Xiaoliang Qi, T. Takayanagi, and A. Wall for useful discussions.
The research of ES is supported by the  National Science Foundation under grant number PHY-1720397, a Simons Foundation Investigator Award, and the Simons Foundation Origins of the Universe program (Modern Inflationary Cosmology collaboration).  GT is supported by CONICET (PIP grant 11220150100299), ANPCYT PICT grant 2015-1224, UNCuyo and CNEA.

%%%%%%%%%%%%%%%%%%%%%%%%%%%%%%%%%%%%%%%%%%%%%%%
%%%%%%%%%%%%%%%%%%%%%%%%%%%%%%%%%%%%%%%%%%%%%%%
%%%%%%%%%%%%%%%%%%%%%%%%%%%%%%%%%%%%%%%%%%%%%%%
\appendix

\section{Interactions between two matter sectors in $dS_{d+1}$}\label{app:mix}

Here we will describe the mixing interaction (\ref{Smix}) between the two CFTs in a complementary way to the transmission coefficient analysis in \cite{Hologravity}.  These analyses are tractable only for sufficiently low operator dimensions $\Delta\ll c$ that the correlation functions respect large-$N$ factorization.  One can compute the generating function $Z_{CFT}[J]$ for each CFT on $dS_d$ using the gravity side as in \cite{HarlowStanford}, with $J$ the source of a scalar operator ${\cal O}$ of dimension $\Delta$.  At the Gaussian level,  this takes the form
\be\label{GformZ}
Z_{CFT, G}[J] = \exp\left( -\int_x\int_y J (x) G_{AdS}(x,y) J(y) \right)
\ee
where $G_{AdS}$ is the 2-point function in the CFT, expressible in terms of $\Gamma$ functions \cite{HarlowStanford}.  In an angular momentum basis for Euclidean $dS_d$, this Green's function behaves like $G_{AdS}\sim\ell^{2\Delta-d}$ at large total angular momentum $\ell$.  Again at $\Delta\sim c$, where the dominant states in our entanglement calculation reside, non-Gaussian effects will not be suppressed; here we are just getting an idea of the strength of the mixing for lower $\Delta$ where the calculation is tractable.

Next, we wish to incorporate interactions of the form (\ref{Smix}).   Consider the generating function $Z[J]$ for each sector, in the presence of the mixing interaction, tracing out the other sector.  Denoting the degrees of freedom of the two CFTs by $\chi$ and $\tilde\chi$ respectively, we can write
\bea\label{ZJ}
Z[J] &=& \int D\chi \int D\tilde\chi e^{i(S_{CFT}(\chi)+\int \lambda {\cal O}\widetilde{\cal O}+\int J{\cal O}+S_{CFT}(\tilde\chi))} \nonumber\\
&=& \int D\chi e^{i(S_{CFT}(\chi)+ \int {\cal O} J)}Z_{CFT}[\lambda {\cal O}] \,.
\eea
At the Gaussian level for the large-$N$ factorized fields, we have (\ref{GformZ}), and correspondingly 
\be\label{GformS}
S_{CFT, G}\sim \int_x\int_y {\cal O}(x) G_{AdS}^{-1}(x, y) {\cal O}(y) \,.
\ee
Plugging this into (\ref{ZJ}) yields
\be\label{ZJdS}
Z_G[J]\sim e^{-\int_x\int_y J(x) (\lambda^2 G_{AdS}+G_{AdS}^{-1}) J(y)}\,.
\ee
We can now solve for $\lambda$ by doing a gravity-side calculation of the 2-point function $G_{dS}=\langle {\cal O}{\cal O}\rangle$ in one warped throat, say the $w<\pi \ell_{dS}/2$ side of (\ref{dSslicing}), and imposing
\be\label{lambdaeq}
G_{dS}= \lambda^2 G_{AdS}+G_{AdS}^{-1} ~~~~\Rightarrow  ~~~~ \lambda = \frac{1}{G_{AdS}}\sqrt{G_{dS}G_{AdS}-1}\,,
\ee
where in the last expression, we work in a diagonal (angular momentum) basis.
So the Gaussian action for ${\cal O}, \widetilde{\cal O}$ is of the form
\be\label{Oac}
\sum \frac{1}{G_{AdS}}\left(  {\cal O}{\cal O}+{\cal O}\widetilde{\cal O} \sqrt{G_{dS}G_{AdS}-1} + \widetilde{\cal O}\widetilde{\cal O}  \right)\,.
\ee
We can determine $G_{dS}$ via a straightforward calculation along the lines of \cite{HarlowStanford}.  The result has the property that $G_{dS}\sim \ell^1$ for $\ell\gg 1$, independently of $\Delta$.  Since we are interested in interpreting the volume-law entropy, let us focus on this $\ell\gg 1$ regime.
Plugging $G_{dS}\sim \ell$, $G_{AdS}\sim \ell ^{2\Delta-d}$ into (\ref{Oac}), we see that the mixing interaction dominates over the pure CFT terms in this effective Gaussian action at large angular momentum $\ell$.  In general, the dS/dS correspondence may require irrelevant operators deforming each CFT separately, which could be incorporated in a similar way at the Gaussian level.

%%%%%%%%%%%%%%%%%%%%%%%%%%%%%%%%%%%%%%%%%%%%%%%
%%%%%%%%%%%%%%%%%%%%%%%%%%%%%%%%%%%%%%%%%%%%%%%
%%%%%%%%%%%%%%%%%%%%%%%%%%%%%%%%%%%%%%%%%%%%%%%
\section{Liouville quantum gravity and $dS_2$}\label{app:liouville}

Here we spell out the role of the Liouville gravity and how it leads to the $dS_2$ evolution.
This is essentially a review of the framework of Liouville quantum cosmology \cite{LiouvilleCosmo, Martinec},  applied to our system.  

We write the metric as $g_{\alpha\beta}=e^{\kappa\phi}\eta_{\alpha\beta}$; in cosmological solutions, $e^{\kappa\phi/2}$ plays the role of the scale factor.  We will be interested in ground states corresponding to a classical global $dS_2$ solution,
\bea\label{dS2}
ds^2_{dS_2} &=& -d\tau^2 +\ell_{dS}^2 \cosh^2(\tau/\ell_{dS}) d\theta^2, ~~~~ \theta = \theta+ 2\pi  \nonumber\\
  &=& e^{\kappa\phi}\left(-d\eta^2+d\sigma^2\right), ~~~~ \sigma = \sigma+2\pi\ell_{dS}  
\eea
with curvature ${\cal R}=2/\ell_{dS}^2$.  The proper volume at the neck ($\tau=0$) is $2\pi \ell_{dS}$, the same as that at the most ultraviolet $dS_2$ slice of the $dS_3$ metric (\ref{dSslicing}).

For simplicity, and since we aim to first understand the state corresponding to unexcited de Sitter spacetime, let us solve the momentum constraint with zero spatial momentum in the matter sector.
The minisuperspace\footnote{It is not necessary to restrict to minisuperspace, but as discussed in \cite{LiouvilleCosmo}\ and elsewhere, this turns out to be a good guide.} Hamiltonian constraint then takes the form
\be\label{Hamconstraint}
\left\{-\frac{\partial^2}{\partial\phi^2}- 2 {\cal H}_{CFT_1} - 2{{\cal H}}_{CFT_2}-2 {\cal H}_{mix}-\dots - \frac{4}{\kappa^2} e^{\kappa\phi} \right\}\Psi = 0\,,
\ee
with $\kappa \simeq 2\sqrt{3/c_{tot}}$ at large total central charge in the infrared of our mixed matter theory, with Hamiltonian 
${\cal H}_{tot}={\cal H}_{CFT_1}+{\cal H}_{CFT_2}+{\cal H}_{mix}+\dots$.  

For the pure de Sitter solution, we are interested in the lowest energy solution in the matter sector; excited energy levels
lead to a deformation up to a point where they cause large backreaction.  
We work on the cylinder, and include the Casimir energy $-c_{tot}/12$ as part of the Hamiltonian of the infrared CFT of the full system:  
the eigenstates of ${\cal H}_{tot}$ are $E_{tot, IR}=\Delta_{tot, IR} - c_{tot}/12$.

When the Liouville theory is coupled to a CFT matter sector, the solution to the Wheeler-DeWitt equation (\ref{Hamconstraint}) behaves as global de Sitter spacetime, or an excitation of it that still bounces, for negative total CFT energy \cite{LiouvilleCosmo, Martinec}.
We can see this as follows.
For this window $E_{tot, IR}<0$, there is a bouncing solution.  
At the minimal energy, with $\Delta_{tot, IR} = 0$, the classical turning point occurs at $\phi_*=0$. More generally, it occurs at 
\be\label{turning}
e^{\kappa\phi_{*}}=1-\frac{12\Delta_{tot, IR}}{c_{tot, IR}}
\ee
leading to a smaller neck for $\Delta_{tot, IR}>0$, and a taller Penrose diagram.  For pure CFTs, at  the upper end of this window, the neck approaches zero proper size according to (\ref{turning}), leading to $\phi_*\to -\infty$.  At energies above this window, the matter backreaction is too great to support the full contracting and expanding phase.  For more involved matter theories, the Liouville field couples nontrivially, leading to richer 2d cosmological dynamics \cite{Martinec}.

In the classically disallowed regime $\phi\to -\infty$, these wavefunctions decay exponentially.  The Hamiltonian constraint leads to solutions $e^{i \hat \omega\phi}$ with imaginary frequency 
\be\label{p0}
\hat \omega =\sqrt{2\(\Delta_{tot, IR} -\frac{c_{tot, IR}}{12}\)}\,.
\ee
In the context of the worldsheet theory of supercritical strings, $X^0 = \phi \sqrt{\alpha'}$ plays the role of the timelike embedding coordinate of the string, with solutions $\propto e^{i \hat p^0 X^0}$ corresponding to spacetime energy levels ${\hat p^0}=\hat\omega/\sqrt{\alpha'}$.
\footnote{As reviewed recently in \cite{dilatontracer}, on the cylinder the spacetime momenta are shifted from those on the plane, which include an additional piece $\propto i \sqrt{c} \phi$.}  These energy levels may be obtained from the $T\bar T$ deformation \cite{ZTTbar, TTbarJT}, potentially eliminating any direct reference to gravity, which would be interesting to pursue in the future.

%%%%%%%%%%%%%%%%%%%%%%%%%%%%%%%%%%%%%%%%%%%%%%%
%%%%%%%%%%%%%%%%%%%%%%%%%%%%%%%%%%%%%%%%%%%%%%%
%%%%%%%%%%%%%%%%%%%%%%%%%%%%%%%%%%%%%%%%%%%%%%%
\section{A lattice model with volume law entanglement}\label{app:spin}

The physics of the $d$ dimensional dual theory involves an interplay between interactions between the two matter sectors, and entanglement between them over the volume $V_{d-1}$.  In this appendix, we describe a lattice model that displays some of the relevant features, and where the calculations can be carried out explicitly.

Let us consider a lattice where at each lattice point we have two types of spins, $\vec S$ and $\vec S'$. These will model the two CFT sectors. There are nearest-neighbor antiferromagnetic couplings, and also on-site couplings between the different spins:
\be\label{eq:latticeH}
H = \sum_{\langle ij\rangle}(J_{ij} \vec S_i \cdot \vec S_j+J'_{ij} \vec S'_i \cdot \vec S'_j)+ \sum_i\, K_i\, \vec S_i \cdot \vec S'_i\,.
\ee
All the couplings $J, J', K$ are positive. This is basically a discretized version of two scalar fields coupled via an interaction of the form $\phi_1^2 \phi_2^2$. Now we take the limit $K \gg J, J'$. In this strong coupling limit, the ground state of the system is that the two spins $\vec S_i, \vec S_i'$ form a singlet (a ``bond'') at each lattice site,
\be\label{eq:Psi}
| \Psi \rangle = \prod_i \frac{1}{\sqrt{2}}\left(| \uparrow \downarrow \rangle_i -|  \downarrow \uparrow \rangle_i \right)\,.
\ee

We now trace out all the $\vec S'_i$, obtaining the reduced density matrix
\be
\rho=\text{Tr}_{S'} (| \Psi \rangle \langle \Psi |)= \prod_i\,\frac{1}{2}(|\uparrow \rangle_i \langle \uparrow|_i+|\downarrow \rangle_i \langle \downarrow |_i)\,.
\ee
We get a maximally mixed state at each site. The von Neumann entropy for the full system is simply
\be
S= -\text{Tr}(\rho\,\log \rho)= N \log 2
\ee
where $N$ is the number of sites. This reproduces as it should the dimension of the Hilbert space, $e^S= 2^N$.

We can also consider an entangling region $\Sigma$, and trace over all the spins $\vec S$ in the complement of the region (after having traced out all the $\vec S'$). Denoting $\rho_\Sigma = \text{Tr}_{\bar \Sigma}\,\rho$, we find the entanglement entropy
\be
S_\Sigma = N_\Sigma \log 2 = \frac{\text{Vol}_\Sigma}{a^{d-1}}\,\log 2\,.
\ee
Here $N_\Sigma = \text{Vol}_\Sigma/a^{d-1}$ is the number of spins inside $\Sigma$, $a$ is the lattice spacing, and the lattice has $d-1$ spatial dimensions. Intuitively, the entanglement entropy counts the number of bonds (spin singlets) between the inside and the outside of $\Sigma$; each bond contributes $\log 2$ to the von Neumann entropy. As $K \to \infty$, each of the spins $\vec S_i$ inside the entangling region $\Sigma$ is in a singlet with $\vec S'_i$. Now, the spins $\vec S'_i$ have been traced out -- they behave like an external bath for the spins $\vec S_i$. Therefore every spin inside $\Sigma$ is connected to a bond outside $\Sigma$, and we obtain the volume law.

This simple example provides a toy model illustrating some of the features of what we are finding in the dS/dS duality. The two QFTs are modeled here by the two types of spins, and the large irrelevant couplings between the sectors are represented by the large on-site coupling $K \vec S \cdot \vec S'$. The fact that the interaction is irrelevant does not seem to be important for the entanglement entropy property; rather, the point is that it dominates over the interactions within each sector. In the lattice model, tracing out one sector completely leads to a volume law for the entanglement entropy associated to finite regions of the lattice. For $K/J$ finite but large, some of the bonds between $\vec S$ and $\vec S'$ would break, and instead there would be bonds between 
spins $\vec S_i$ at different sites. This leads to an area term correction.

%%%%%%%%%%%%%%%%%%%%%%%%%%%%%%%%%%%%%%%%%%%%%%%
%%%%%%%%%%%%%%%%%%%%%%%%%%%%%%%%%%%%%%%%%%%%%%%
\section{Holographic entanglement for subregions in dS/dS}\label{app:EE}

In this appendix we expand upon the calculation of different entanglement measures using holography. We focus on the von Neumann and Renyi entropies, as well as the mutual information. We will exhibit a volume law, as well as vanishing mutual information.

As discussed in the main text in \S\ref{sec:volumecomment}, the holographic entanglement entropy has a clear interpretation only for a region that covers the whole space $S^{d-1}$ of global $dS_d$. This arises from tracing out completely one of the two matter sectors, and corresponds to (\ref{SVN}, \ref{Renyidef}). Nevertheless, in this appendix we will perform a more general RT calculation for regions of size smaller than $\ell_{dS}$, and will also obtain a volume law. As the entangling region approaches the whole space, we will recover (\ref{VolEnt}). The physical interpretation of these entropies for smaller regions is not completely clear, but this could provide an intriguing clue regarding locality at scales smaller than $\ell_{dS}$.

Consider an entangling region $\Sigma$ in the dual $d$-dimensional theory, at $\sin(w/\ell_{dS})=1$ and on the $S^{d-1}$ at $\tau=0$ in global de Sitter (\ref{dSslicing}). We allow the extremal surface $\mathcal M$ to extend into the bulk (meaning $w\ne\pi\ell_{dS}/2$), parameterized by $w=w(y^i)$, where $y^i$ denote the coordinates of the $S^{d-1}$. We will also denote the unit sphere metric by $\t g_{ij}$.  Defining $\hat w=w/\ell_{dS}$,
the metric on this surface is
\be
ds|_{\mathcal M}^2= \ell_{dS}^2 (\partial_i \hat w \, \partial_j \hat w+\sin^2\hat w\,\t g_{ij}) dy^i dy^j\,,
\ee
and the area functional is
\be\label{Aftnal}
A=\ell_{dS}^{d-1} \int d^{d-1}y\,\text{det}^{1/2}(\partial_i \hat w \, \partial_j \hat w+\sin^2 \hat w\,\t g_{ij})\,.
\ee
We find the extremal surface by varying this with respect to $w(y_i)$.  For example, in the $d=2$ case this generates the geodesic equation, whose solution lies along the great circle at
\be
\hat w(y^i) = \frac{\pi}{2}\,.
\ee
A similar result holds for higher dimensions. The variation of (\ref{Aftnal}) with respect to $\hat w(y_i)$ has two types of terms; one is proportional to the variation of the warp factor, $\sin \hat w \cos\hat w$, which vanishes on the UV slice.  The remaining terms contain derivatives $\partial_i \hat w$, so since the other contribution vanishes, a consistent solution is the constant one, $\hat w(y_i)=\pi/2$.

This gives a volume law for the entanglement entropy,
\be\label{eq:volume}
{\cal S}(\Sigma) = \frac{V_\Sigma}{4 G_{d+1}}\,.
\ee
This volume law holds for regions $\Sigma$ of size smaller than $\ell_{dS}$. As the entangling region covers the whole space, we recover (\ref{VolEnt}).

It is not difficult to understand this volume law from the maximal mixing of the density matrix $\rho_1$ for QFT$_1$ as we argued in the main text.  Partially tracing $\rho_1$ over any subsystem produces a reduced density matrix $\rho_\Sigma$ that is maximally mixed, and the von Neumann entropy of a maximally mixed $\rho_\Sigma$ satisfies a volume law.

The maximal mixing of $\rho_\Sigma$ also ensures that its Renyi entropies ${\cal S}_n(\Sigma)$ must be independent of $n$ and equal to the von Neumann entropy (\ref{eq:volume}).  In principle, this can also be obtained directly from the area of cosmic branes, although their backreaction is nontrivial to calculate in this case.

We now compute the mutual information between two disjoint regions $A$ and $B$ in QFT$_1$:
\be
I(A, B) = {\cal S}(A)+{\cal S}(B)-{\cal S}(A \cup B)\,.
\ee
This measures the correlations between the two regions~\cite{wolf2008area}. Since the extremal surfaces always stay at the UV slice, the sum of their areas equals the area of the union, and the mutual information vanishes. This means that at large $c$ there are no correlations between disjoint subsystems of QFT$_1$ except for a small number of degrees of freedom (as we expect a subleading nonzero correction from quantum effects in the bulk).

The vanishing of the mutual information between $A$ and $B$ in QFT$_1$ can be easily understood as a consequence of QFT$_1$ being maximally entangled with QFT$_2$ (to the leading order in the large-$c$ limit).  It essentially follows from the monogamy of quantum entanglement.  The region $A$ is maximally entangled with some subsystem $C$ of QFT$_2$, which prevents $A$ from being correlated with $B$ at all.  In other words, strong subadditivity together with the pureness of $A \cup C$ requires
\be
{\cal S}(A \cup B) = {\cal S}(A \cup B) + {\cal S}(A \cup C) \geq {\cal S}(A) + {\cal S}(A\cup B\cup C) = {\cal S}(A) + {\cal S}(B)
\ee
which immediately leads to a vanishing mutual information.

Let us compare this with the mutual information in CFTs with AdS dual. There are two possible solutions for the extremal surface of $A \cup B$. One is a union of separate extremal surfaces of $A$ and $B$. There is also another solution, with a tube that connects $A$ and $B$ going into the bulk. For sufficiently separated regions, the first solution dominates, and the mutual information vanishes. However, as the regions approach each other, there is a phase transition and the tube-like solution becomes the dominant one, giving a nonzero mutual information~\cite{Headrick:2010zt}. This is suggestive of a Hagedorn transition.\footnote{We thank H. Casini for discussions on this point. See also~\cite{Klebanov:2007ws}.} In contrast, in the de Sitter case, if we consider regions smaller than the full neck, we find that this transition is absent -- the mutual information always vanishes to the leading order in $1/c$.


\begin{thebibliography}{99}

\bibitem{RTHRT}

S.~Ryu and T.~Takayanagi,
  ``Holographic derivation of entanglement entropy from AdS/CFT,''
  Phys.\ Rev.\ Lett.\  {\bf 96}, 181602 (2006)
  %doi:10.1103/PhysRevLett.96.181602
  [hep-th/0603001].
  %%CITATION = doi:10.1103/PhysRevLett.96.181602;%%
  %1506 citations counted in INSPIRE as of 02 Apr 2018
  
V.~E.~Hubeny, M.~Rangamani and T.~Takayanagi,
  ``A Covariant holographic entanglement entropy proposal,''
  JHEP {\bf 0707}, 062 (2007)
  %doi:10.1088/1126-6708/2007/07/062
  [arXiv:0705.0016 [hep-th]].
  %%CITATION = doi:10.1088/1126-6708/2007/07/062;%%
  %593 citations counted in INSPIRE as of 02 Apr 2018  

%\cite{Dong:2012afa}

\bibitem{Xietal}
  X.~Dong,
  ``Holographic Entanglement Entropy for General Higher Derivative Gravity,''
  JHEP {\bf 1401}, 044 (2014)
  %doi:10.1007/JHEP01(2014)044
  [arXiv:1310.5713 [hep-th]].
  %%CITATION = doi:10.1007/JHEP01(2014)044;%%
  %179 citations counted in INSPIRE as of 17 Apr 2018

J.~Camps,
  ``Generalized entropy and higher derivative Gravity,''
  JHEP {\bf 1403}, 070 (2014)
  %doi:10.1007/JHEP03(2014)070
  [arXiv:1310.6659 [hep-th]].
  %%CITATION = doi:10.1007/JHEP03(2014)070;%%
  %137 citations counted in INSPIRE as of 17 Apr 2018

\bibitem{JuanTwoSided}
J.~M.~Maldacena,
  ``Eternal black holes in anti-de Sitter,''
  JHEP {\bf 0304}, 021 (2003)
  %doi:10.1088/1126-6708/2003/04/021
  [hep-th/0106112].
  %%CITATION = doi:10.1088/1126-6708/2003/04/021;%%
  %637 citations counted in INSPIRE as of 02 Apr 2018
  
\bibitem{VanR}
 M.~Van Raamsdonk,
  ``Comments on quantum gravity and entanglement,''
  arXiv:0907.2939 [hep-th].
  %%CITATION = ARXIV:0907.2939;%%
  %171 citations counted in INSPIRE as of 02 Apr 2018
  
\bibitem{EREPR}  
J.~Maldacena and L.~Susskind,
  ``Cool horizons for entangled black holes,''
  Fortsch.\ Phys.\  {\bf 61}, 781 (2013)
  %doi:10.1002/prop.201300020
  [arXiv:1306.0533 [hep-th]].
  %%CITATION = doi:10.1002/prop.201300020;%%
  %420 citations counted in INSPIRE as of 02 Apr 2018  
  

\bibitem{ShenkerStanford}

 S.~H.~Shenker and D.~Stanford,
  ``Multiple Shocks,''
  JHEP {\bf 1412}, 046 (2014)
  %doi:10.1007/JHEP12(2014)046
  [arXiv:1312.3296 [hep-th]].
  %%CITATION = doi:10.1007/JHEP12(2014)046;%%
  %143 citations counted in INSPIRE as of 07 Apr 2018

 \bibitem{Traversable}
 
  P.~Gao, D.~L.~Jafferis and A.~Wall,
  ``Traversable Wormholes via a Double Trace Deformation,''
  JHEP {\bf 1712}, 151 (2017)
  %doi:10.1007/JHEP12(2017)151
  [arXiv:1608.05687 [hep-th]].
  %%CITATION = doi:10.1007/JHEP12(2017)151;%%
  %26 citations counted in INSPIRE as of 02 Apr 2018 
  
  J.~Maldacena, D.~Stanford and Z.~Yang,
  ``Diving into traversable wormholes,''
  Fortsch.\ Phys.\  {\bf 65}, no. 5, 1700034 (2017)
  %doi:10.1002/prop.201700034
  [arXiv:1704.05333 [hep-th]].
  %%CITATION = doi:10.1002/prop.201700034;%%
  %22 citations counted in INSPIRE as of 02 Apr 2018
  
  J.~Maldacena and X.~L.~Qi,
  ``Eternal traversable wormhole,''
  arXiv:1804.00491 [hep-th].
  %%CITATION = ARXIV:1804.00491;%%

\bibitem{MarkVradial}

V.~Balasubramanian, M.~B.~McDermott and M.~Van Raamsdonk,
  ``Momentum-space entanglement and renormalization in quantum field theory,''
  Phys.\ Rev.\ D {\bf 86}, 045014 (2012)
  %doi:10.1103/PhysRevD.86.045014
  [arXiv:1108.3568 [hep-th]].
  %%CITATION = doi:10.1103/PhysRevD.86.045014;%%
  %69 citations counted in INSPIRE as of 02 Apr 2018
  
\bibitem{sphereexample}

A.~Mollabashi, N.~Shiba and T.~Takayanagi,
  ``Entanglement between Two Interacting CFTs and Generalized Holographic Entanglement Entropy,''
  JHEP {\bf 1404}, 185 (2014)
  %doi:10.1007/JHEP04(2014)185
  [arXiv:1403.1393 [hep-th]].
  %%CITATION = doi:10.1007/JHEP04(2014)185;%%
  %25 citations counted in INSPIRE as of 12 Apr 2018

  M.~R.~Mohammadi Mozaffar and A.~Mollabashi,
  ``On the Entanglement Between Interacting Scalar Field Theories,''
  JHEP {\bf 1603}, 015 (2016)
  %doi:10.1007/JHEP03(2016)015
  [arXiv:1509.03829 [hep-th]].
  %%CITATION = doi:10.1007/JHEP03(2016)015;%%
  %8 citations counted in INSPIRE as of 08 Jun 2018

  M.~R.~Mohammadi Mozaffar and A.~Mollabashi,
  ``Entanglement in Lifshitz-type Quantum Field Theories,''
  JHEP {\bf 1707}, 120 (2017)
  %doi:10.1007/JHEP07(2017)120
  [arXiv:1705.00483 [hep-th]].
  %%CITATION = doi:10.1007/JHEP07(2017)120;%%
  %10 citations counted in INSPIRE as of 08 Jun 2018

 A.~Karch and C.~F.~Uhlemann,
  ``Holographic entanglement entropy and the internal space,''
  Phys.\ Rev.\ D {\bf 91}, no. 8, 086005 (2015)
  %doi:10.1103/PhysRevD.91.086005
  [arXiv:1501.00003 [hep-th]].
  %%CITATION = doi:10.1103/PhysRevD.91.086005;%%
  %8 citations counted in INSPIRE as of 12 Apr 2018

\bibitem{GibbonsHawking}

G.~W.~Gibbons and S.~W.~Hawking,
  ``Cosmological Event Horizons, Thermodynamics, and Particle Creation,''
  Phys.\ Rev.\ D {\bf 15}, 2738 (1977).
  %doi:10.1103/PhysRevD.15.2738
  %%CITATION = doi:10.1103/PhysRevD.15.2738;%%
  %1957 citations counted in INSPIRE as of 07 Apr 2018
 
  \bibitem{Micromanaging}
 
  X.~Dong, B.~Horn, E.~Silverstein and G.~Torroba,
  ``Micromanaging de Sitter holography,''
  Class.\ Quant.\ Grav.\  {\bf 27}, 245020 (2010)
  %doi:10.1088/0264-9381/27/24/245020
  [arXiv:1005.5403 [hep-th]].
  %%CITATION = doi:10.1088/0264-9381/27/24/245020;%%
  %52 citations counted in INSPIRE as of 05 Mar 2018


\bibitem{dSCFT}

A.~Strominger,
  ``The dS / CFT correspondence,''
  JHEP {\bf 0110}, 034 (2001)
  %doi:10.1088/1126-6708/2001/10/034
  [hep-th/0106113].
  %%CITATION = doi:10.1088/1126-6708/2001/10/034;%%
  %804 citations counted in INSPIRE as of 12 Apr 2018
  
D.~Anninos, T.~Hartman and A.~Strominger,
  ``Higher Spin Realization of the dS/CFT Correspondence,''
  Class.\ Quant.\ Grav.\  {\bf 34}, no. 1, 015009 (2017)
  %doi:10.1088/1361-6382/34/1/015009
  [arXiv:1108.5735 [hep-th]].
  %%CITATION = doi:10.1088/1361-6382/34/1/015009;%%
  %167 citations counted in INSPIRE as of 17 Apr 2018
  
\bibitem{Hologravity} 

%\cite{Alishahiha:2005dj}
%\bibitem{Alishahiha:2005dj} 
  
  M.~Alishahiha, A.~Karch and E.~Silverstein,
  ``Hologravity,''
  JHEP {\bf 0506}, 028 (2005)
  %doi:10.1088/1126-6708/2005/06/028
  [hep-th/0504056].
  %%CITATION = doi:10.1088/1126-6708/2005/06/028;%%
  %37 citations counted in INSPIRE as of 05 Mar 2018
%\cite{Alishahiha:2004md}

%\bibitem{Alishahiha:2004md} 
  M.~Alishahiha, A.~Karch, E.~Silverstein and D.~Tong,
  ``The dS/dS correspondence,''
  AIP Conf.\ Proc.\  {\bf 743}, 393 (2005)
  %doi:10.1063/1.1848341
  [hep-th/0407125].
  %%CITATION = doi:10.1063/1.1848341;%%
  %87 citations counted in INSPIRE as of 05 Mar 2018


 \bibitem{FRWFRW}
   X.~Dong, B.~Horn, S.~Matsuura, E.~Silverstein and G.~Torroba,
  ``FRW solutions and holography from uplifted AdS/CFT,''
  Phys.\ Rev.\ D {\bf 85}, 104035 (2012)
  %doi:10.1103/PhysRevD.85.104035
  [arXiv:1108.5732 [hep-th]].
  %%CITATION = doi:10.1103/PhysRevD.85.104035;%%
  %26 citations counted in INSPIRE as of 05 Mar 2018 
  

\bibitem{FRWCFT}

 B.~Freivogel, Y.~Sekino, L.~Susskind and C.~P.~Yeh,
  ``A Holographic framework for eternal inflation,''
  Phys.\ Rev.\ D {\bf 74}, 086003 (2006)
  %doi:10.1103/PhysRevD.74.086003
  [hep-th/0606204].
  %%CITATION = doi:10.1103/PhysRevD.74.086003;%%
  %94 citations counted in INSPIRE as of 12 Apr 2018

\bibitem{TASIcosmo}

 E.~Silverstein,
  ``TASI lectures on cosmological observables and string theory,''
  $doi:10.1142/9789813149441_0009$
  arXiv:1606.03640 [hep-th].
  %%CITATION = doi:10.1142/9789813149441_0009;%%
  %9 citations counted in INSPIRE as of 03 Apr 2018
  
\bibitem{musings}

D.~Anninos,
  ``De Sitter Musings,''
  Int.\ J.\ Mod.\ Phys.\ A {\bf 27}, 1230013 (2012)
  %doi:10.1142/S0217751X1230013X
  [arXiv:1205.3855 [hep-th]].
  %%CITATION = doi:10.1142/S0217751X1230013X;%%
  %57 citations counted in INSPIRE as of 12 Apr 2018  

\bibitem{RS}
L.~Randall and R.~Sundrum,
  ``An Alternative to compactification,''
  Phys.\ Rev.\ Lett.\  {\bf 83}, 4690 (1999)
  %doi:10.1103/PhysRevLett.83.4690
  [hep-th/9906064].
  %%CITATION = doi:10.1103/PhysRevLett.83.4690;%%
  %6154 citations counted in INSPIRE as of 05 Apr 2018

\bibitem{HoloRG} 
  X.~Dong, B.~Horn, E.~Silverstein and G.~Torroba,
  ``Moduli Stabilization and the Holographic RG for AdS and dS,''
  JHEP {\bf 1306}, 089 (2013)
  %doi:10.1007/JHEP06(2013)089
  [arXiv:1209.5392 [hep-th]].
  %%CITATION = doi:10.1007/JHEP06(2013)089;%%
  %4 citations counted in INSPIRE as of 05 Mar 2018

\bibitem{HarlowStanford}

D.~Harlow and D.~Stanford,
  ``Operator Dictionaries and Wave Functions in AdS/CFT and dS/CFT,''
  arXiv:1104.2621 [hep-th].
  %%CITATION = ARXIV:1104.2621;%%
  %112 citations counted in INSPIRE as of 03 Apr 2018

J.~M.~Maldacena,
  ``Non-Gaussian features of primordial fluctuations in single field inflationary models,''
  JHEP {\bf 0305}, 013 (2003)
  %doi:10.1088/1126-6708/2003/05/013
  [astro-ph/0210603].
  %%CITATION = doi:10.1088/1126-6708/2003/05/013;%%
  %1750 citations counted in INSPIRE as of 03 Apr 2018

\bibitem{ZTTbar}

A.~B.~Zamolodchikov,
  ``Expectation value of composite field T anti-T in two-dimensional quantum field theory,''
  hep-th/0401146.
  %%CITATION = HEP-TH/0401146;%%
  %24 citations counted in INSPIRE as of 03 Apr 2018

\bibitem{TTbarJT}

S.~Dubovsky, V.~Gorbenko and M.~Mirbabayi,
  ``Asymptotic fragility, near AdS$_{2}$ holography and $ T\overline{T} $,''
  JHEP {\bf 1709}, 136 (2017)
  %doi:10.1007/JHEP09(2017)136
  [arXiv:1706.06604 [hep-th]].
  %%CITATION = doi:10.1007/JHEP09(2017)136;%%
  %9 citations counted in INSPIRE as of 05 Mar 2018
  
\bibitem{AkiBill}

 W.~Cottrell and A.~Hashimoto,
  ``Comments on $T \bar T$ double trace deformations and boundary conditions,''
  arXiv:1801.09708 [hep-th].
  %%CITATION = ARXIV:1801.09708;%%
  %2 citations counted in INSPIRE as of 05 Mar 2018

\bibitem{Cardy}
J.~Cardy,
  ``The $T\overline T$ deformation of quantum field theory as a stochastic process,''
  arXiv:1801.06895 [hep-th].
  %%CITATION = ARXIV:1801.06895;%%
  %3 citations counted in INSPIRE as of 05 Mar 2018

\bibitem{MarolfKraus}

P.~Kraus, J.~Liu and D.~Marolf,
  ``Cutoff AdS$_3$ versus the $T\bar{T}$ deformation,''
  arXiv:1801.02714 [hep-th].
  %%CITATION = ARXIV:1801.02714;%%
  %5 citations counted in INSPIRE as of 03 Apr 2018
  
 
 \bibitem{Verlinde}
 
  L.~McGough, M.~Mezei and H.~Verlinde,
  ``Moving the CFT into the bulk with $T\bar T$,''
  arXiv:1611.03470 [hep-th].
  %%CITATION = ARXIV:1611.03470;%%
  %22 citations counted in INSPIRE as of 03 Apr 2018

\bibitem{Tunneling}

 S.~Dimopoulos, S.~Kachru, N.~Kaloper, A.~E.~Lawrence and E.~Silverstein,
  ``Generating small numbers by tunneling in multithroat compactifications,''
  Int.\ J.\ Mod.\ Phys.\ A {\bf 19}, 2657 (2004)
  %doi:10.1142/S0217751X04018075
  [hep-th/0106128].
  %%CITATION = doi:10.1142/S0217751X04018075;%%
  %56 citations counted in INSPIRE as of 03 Apr 2018
%\cite{Dimopoulos:2001ui}
%\bibitem{Dimopoulos:2001ui} 
  
  S.~Dimopoulos, S.~Kachru, N.~Kaloper, A.~E.~Lawrence and E.~Silverstein,
  ``Small numbers from tunneling between brane throats,''
  Phys.\ Rev.\ D {\bf 64}, 121702 (2001)
  %doi:10.1103/PhysRevD.64.121702
  [hep-th/0104239].
  %%CITATION = doi:10.1103/PhysRevD.64.121702;%%
  %68 citations counted in INSPIRE as of 03 Apr 2018


  
\bibitem{CFTspectrum}

 T.~Hartman, C.~A.~Keller and B.~Stoica,
 ``Universal Spectrum of 2d Conformal Field Theory in the Large c Limit,''
  JHEP {\bf 1409}, 118 (2014)
  %doi:10.1007/JHEP09(2014)118
  [arXiv:1405.5137 [hep-th]].
  %%CITATION = doi:10.1007/JHEP09(2014)118;%%
  %108 citations counted in INSPIRE as of 04 Apr 2018  
  
 C.~A.~Keller and A.~Maloney,
  ``Poincare Series, 3D Gravity and CFT Spectroscopy,''
  JHEP {\bf 1502}, 080 (2015)
  %doi:10.1007/JHEP02(2015)080
  [arXiv:1407.6008 [hep-th]].
  %%CITATION = doi:10.1007/JHEP02(2015)080;%%
  %28 citations counted in INSPIRE as of 17 Apr 2018 

S.~Hellerman,
  ``A Universal Inequality for CFT and Quantum Gravity,''
  JHEP {\bf 1108}, 130 (2011)
  %doi:10.1007/JHEP08(2011)130
  [arXiv:0902.2790 [hep-th]].
  %%CITATION = doi:10.1007/JHEP08(2011)130;%%
  %82 citations counted in INSPIRE as of 07 Apr 2018

  
  N.~Benjamin, M.~C.~N.~Cheng, S.~Kachru, G.~W.~Moore and N.~M.~Paquette,
  ``Elliptic Genera and 3d Gravity,''
  Annales Henri Poincare {\bf 17}, no. 10, 2623 (2016)
  %doi:10.1007/s00023-016-0469-6
  [arXiv:1503.04800 [hep-th]].
  %%CITATION = doi:10.1007/s00023-016-0469-6;%%
  %26 citations counted in INSPIRE as of 07 Apr 2018

  
\bibitem{Nbound}

 T.~Banks,
  ``Cosmological breaking of supersymmetry?,''
  Int.\ J.\ Mod.\ Phys.\ A {\bf 16}, 910 (2001)
  %doi:10.1142/S0217751X01003998
  [hep-th/0007146].
  %%CITATION = doi:10.1142/S0217751X01003998;%%
  %339 citations counted in INSPIRE as of 17 Apr 2018

 R.~Bousso,
  ``Positive vacuum energy and the N bound,''
  JHEP {\bf 0011}, 038 (2000)
  %doi:10.1088/1126-6708/2000/11/038
  [hep-th/0010252].
  %%CITATION = doi:10.1088/1126-6708/2000/11/038;%%
  %192 citations counted in INSPIRE as of 17 Apr 2018
  
\bibitem{Narayan:2017xca} 
  K.~Narayan,
  ``On extremal surfaces and de Sitter entropy,''
  Phys.\ Lett.\ B {\bf 779}, 214 (2018)
  %doi:10.1016/j.physletb.2018.02.010
  [arXiv:1711.01107 [hep-th]].
  %%CITATION = doi:10.1016/j.physletb.2018.02.010;%%

\bibitem{Lewkowycz:2013nqa} 
  A.~Lewkowycz and J.~Maldacena,
 ``Generalized gravitational entropy,''
  JHEP {\bf 1308}, 090 (2013)
  %doi:10.1007/JHEP08(2013)090
  [arXiv:1304.4926 [hep-th]].
  %%CITATION = doi:10.1007/JHEP08(2013)090;%%
  %368 citations counted in INSPIRE as of 13 Apr 2018

\bibitem{Dong:2016fnf} 
  X.~Dong,
  ``The Gravity Dual of Renyi Entropy,''
  Nature Commun.\  {\bf 7}, 12472 (2016)
  %doi:10.1038/ncomms12472
  [arXiv:1601.06788 [hep-th]].
  %%CITATION = doi:10.1038/ncomms12472;%%
  %40 citations counted in INSPIRE as of 13 Apr 2018

\bibitem{Nomura:2017fyh} 
  Y.~Nomura, P.~Rath and N.~Salzetta,
  ``Spacetime from Unentanglement,''
  Phys.\ Rev.\ D {\bf 97}, no. 10, 106010 (2018)
  %doi:10.1103/PhysRevD.97.106010
  [arXiv:1711.05263 [hep-th]].
  %%CITATION = doi:10.1103/PhysRevD.97.106010;%%
  %4 citations counted in INSPIRE as of 08 Jun 2018

\bibitem{SusskindWitten}

 L.~Susskind and E.~Witten,
  ``The Holographic bound in anti-de Sitter space,''
  hep-th/9805114.
  %%CITATION = HEP-TH/9805114;%%
  %706 citations counted in INSPIRE as of 15 Apr 2018

\bibitem{dSCFTcomplex}

M.~Miyaji and T.~Takayanagi,
  ``Surface/State Correspondence as a Generalized Holography,''
  PTEP {\bf 2015}, no. 7, 073B03 (2015)
  %doi:10.1093/ptep/ptv089
  [arXiv:1503.03542 [hep-th]].
  %%CITATION = doi:10.1093/ptep/ptv089;%%
  %41 citations counted in INSPIRE as of 20 Apr 2018
  
Y.~Sato,
  ``Comments on Entanglement Entropy in the dS/CFT Correspondence,''
  Phys.\ Rev.\ D {\bf 91}, no. 8, 086009 (2015)
  %doi:10.1103/PhysRevD.91.086009
  [arXiv:1501.04903 [hep-th]].
  %%CITATION = doi:10.1103/PhysRevD.91.086009;%%
  %4 citations counted in INSPIRE as of 20 Apr 2018

K.~Narayan,
  ``Extremal surfaces in de Sitter spacetime,''
  Phys.\ Rev.\ D {\bf 91}, no. 12, 126011 (2015)
  %doi:10.1103/PhysRevD.91.126011
  [arXiv:1501.03019 [hep-th]].
  %%CITATION = doi:10.1103/PhysRevD.91.126011;%%
  %10 citations counted in INSPIRE as of 20 Apr 2018

\bibitem{ComplexHRT}  

S.~Fischetti and D.~Marolf,
  ``Complex Entangling Surfaces for AdS and Lifshitz Black Holes?,''
  Class.\ Quant.\ Grav.\  {\bf 31}, no. 21, 214005 (2014)
  %doi:10.1088/0264-9381/31/21/214005
  [arXiv:1407.2900 [hep-th]].
  %%CITATION = doi:10.1088/0264-9381/31/21/214005;%%
  %14 citations counted in INSPIRE as of 19 Apr 2018


 \bibitem{WittenHP}
 
 E.~Witten,
  ``Anti-de Sitter space, thermal phase transition, and confinement in gauge theories,''
  Adv.\ Theor.\ Math.\ Phys.\  {\bf 2}, 505 (1998)
  %doi:10.4310/ATMP.1998.v2.n3.a3
  [hep-th/9803131].
  %%CITATION = doi:10.4310/ATMP.1998.v2.n3.a3;%%
  %2689 citations counted in INSPIRE as of 13 Apr 2018

%\cite{Headrick:2010zt}
\bibitem{Headrick:2010zt} 
  M.~Headrick,
  ``Entanglement Renyi entropies in holographic theories,''
  Phys.\ Rev.\ D {\bf 82}, 126010 (2010)
  %doi:10.1103/PhysRevD.82.126010
  [arXiv:1006.0047 [hep-th]].
  %%CITATION = doi:10.1103/PhysRevD.82.126010;%%
  %310 citations counted in INSPIRE as of 13 Apr 2018

%\cite{Pastawski:2015qua}
\bibitem{Pastawski:2015qua} 
  F.~Pastawski, B.~Yoshida, D.~Harlow and J.~Preskill,
  ``Holographic quantum error-correcting codes: Toy models for the bulk/boundary correspondence,''
  JHEP {\bf 1506}, 149 (2015)
  %doi:10.1007/JHEP06(2015)149
  [arXiv:1503.06237 [hep-th]].
  %%CITATION = doi:10.1007/JHEP06(2015)149;%%
  %159 citations counted in INSPIRE as of 19 Apr 2018

%\cite{Hayden:2016cfa}
\bibitem{Hayden:2016cfa} 
  P.~Hayden, S.~Nezami, X.~L.~Qi, N.~Thomas, M.~Walter and Z.~Yang,
  ``Holographic duality from random tensor networks,''
  JHEP {\bf 1611}, 009 (2016)
  %doi:10.1007/JHEP11(2016)009
  [arXiv:1601.01694 [hep-th]].
  %%CITATION = doi:10.1007/JHEP11(2016)009;%%
  %91 citations counted in INSPIRE as of 19 Apr 2018

%\cite{Almheiri:2014lwa}
\bibitem{Almheiri:2014lwa} 
  A.~Almheiri, X.~Dong and D.~Harlow,
  ``Bulk Locality and Quantum Error Correction in AdS/CFT,''
  JHEP {\bf 1504}, 163 (2015)
  %doi:10.1007/JHEP04(2015)163
  [arXiv:1411.7041 [hep-th]].
  %%CITATION = doi:10.1007/JHEP04(2015)163;%%
  %142 citations counted in INSPIRE as of 19 Apr 2018

\bibitem{DHM} 
  X.~Dong, D.~Harlow and D.~Marolf,
  to appear.

\bibitem{BoussoBound}
 R.~Bousso,
  ``A Covariant entropy conjecture,''
  JHEP {\bf 9907}, 004 (1999)
  %doi:10.1088/1126-6708/1999/07/004
  [hep-th/9905177].
  %%CITATION = doi:10.1088/1126-6708/1999/07/004;%%
  %622 citations counted in INSPIRE as of 20 Apr 2018

\bibitem{LiouvilleCosmo}
J.~Polchinski,
  ``A Two-Dimensional Model for Quantum Gravity,''
  Nucl.\ Phys.\ B {\bf 324}, 123 (1989).
  %doi:10.1016/0550-3213(89)90184-3
  %%CITATION = doi:10.1016/0550-3213(89)90184-3;%%
  %196 citations counted in INSPIRE as of 05 Apr 2018

 A.~R.~Cooper, L.~Susskind and L.~Thorlacius,
  ``Two-dimensional quantum cosmology,''
  Nucl.\ Phys.\ B {\bf 363}, 132 (1991).
  %doi:10.1016/0550-3213(91)90238-S
  %%CITATION = doi:10.1016/0550-3213(91)90238-S;%%
  %55 citations counted in INSPIRE as of 05 Apr 2018

\bibitem{Martinec}

B.~Carneiro da Cunha and E.~J.~Martinec,
  ``Closed string tachyon condensation and world sheet inflation,''
  Phys.\ Rev.\ D {\bf 68}, 063502 (2003)
  %doi:10.1103/PhysRevD.68.063502
  [hep-th/0303087].
  %%CITATION = doi:10.1103/PhysRevD.68.063502;%%
  %41 citations counted in INSPIRE as of 03 Apr 2018
  
\bibitem{dilatontracer}

M.~Dodelson, E.~Silverstein and G.~Torroba,
  ``Varying dilaton as a tracer of classical string interactions,''
  Phys.\ Rev.\ D {\bf 96}, no. 6, 066011 (2017)
  %doi:10.1103/PhysRevD.96.066011
  [arXiv:1704.02625 [hep-th]].
  %%CITATION = doi:10.1103/PhysRevD.96.066011;%%
  %3 citations counted in INSPIRE as of 03 Apr 2018

\bibitem{wolf2008area} 
J. Cirac, M. Hastins, F. Verstraete and M. Wolf, ``Area laws in quantum systems: mutual information and correlations,'' Phys. Rev. Lett., {\bf 100}, 7, 070502 (2008) [arXiv:0704.3906].

%\cite{Klebanov:2007ws}
\bibitem{Klebanov:2007ws} 
  I.~R.~Klebanov, D.~Kutasov and A.~Murugan,
  ``Entanglement as a probe of confinement,''
  Nucl.\ Phys.\ B {\bf 796}, 274 (2008)
  %doi:10.1016/j.nuclphysb.2007.12.017
  [arXiv:0709.2140 [hep-th]].
  %%CITATION = doi:10.1016/j.nuclphysb.2007.12.017;%%
  %244 citations counted in INSPIRE as of 13 Apr 2018

\end{thebibliography}
\end{document}